\documentclass{ws-ijmpa}

\begin{document}

\markboth{D. J. Summers}
{Muon Acceleration using FFAG Rings}

%%%%%%%%%%%%%%%%%%%%% Publisher's Area please ignore %%%%%%%%%%%%%%%
%
\catchline{}{}{}{}{}
%
%%%%%%%%%%%%%%%%%%%%%%%%%%%%%%%%%%%%%%%%%%%%%%%%%%%%%%%%%%%%%%%%%%%%

\title{MUON ACCELERATION USING FIXED FIELD, ALTERNATING GRADIENT (FFAG) RINGS}

\author{\footnotesize D. J. SUMMERS\footnote{summers@phy.olemiss.edu
\quad Supported by DE--FG02--91ER40622.}}

\address{Department of Physics and Astronomy, 
University of Mississippi--Oxford\\
University, MS 38677,
USA}

\maketitle

\pub{Received 23 November 2004}{}
%{Revised (Day Month Year)}

\begin{abstract}
Given their 2.2 $\mu$S lifetime, muons must be
accelerated fairly rapidly for a neutrino factory or
muon collider. Muon bunches tend to be large.
Progress in fixed field, alternating gradient
(FFAG) lattices to meet this challenge is reviewed.
FFAG magnets are naturally wide; low momentum muons
move from the low field side of a gradient magnet to the high
field side as they gain energy. This can be exploited to do
double duty and allow a large beam admittance without unduly
increasing the magnetic field volume.  If the amount of RF must be
reduced to optimize cost, an FFAG ring can accommodate extra
orbits.
I describe 
both scaling FFAGs in which the bends in each magnet are energy independent 
and non-scaling FFAGs in which the bends in each magnet do vary with 
muon energy.
In all FFAG designs the sum of the bends in groups of magnets are constant;
otherwise orbits would not close. 
Ways of keeping the 
accelerating beam in phase with the RF are described.
Finally, a 1 MeV {\it proof of principle} scaling FFAG has been built at KEK 
and began accelerating protons in June 2000 with a 1 kHz repetition rate.

\keywords{accelerator; muon; neutrino; black hole.}
\end{abstract}

\section{Introduction}	%) A SECTION HEADING

Scaling FFAG rings were proposed independently a half century ago by 
Ohkawa,$^1$ Symon,$^2$ and Kolomensky.$^3$ The Mid-Western Universities
Research Association (MURA) built both radial-sector (1957) and spiral-sector
(1960) models
and tested them with electrons. However, the serious development of FFAGs
ceased with the ascendancy  of ramping synchrotrons, which allowed smaller 
diameter, smaller bore rings for a given energy and magnetic field.  
Because the voltage
needed to quickly ramp synchrotrons$\,^4$ filled with wide bunches of 
low energy muons is
rather large, FFAGs have recently experienced a renaissance.$^{5,6}$    
The FFAG design permits multiple passages of muons though both RF cavities 
and magnet arcs for reduced cost.

  One reason FFAG rings are of interest today is because they offer
economical muon acceleration for a neutrino factory$\,^{7,8}$ or a 
muon collider.$^9$
At a neutrino factory accelerated muons are stored in a
racetrack to produce neutrino beams
($\mu^- \to e^- \, {\overline{\nu}}_e \, \nu_{\mu}$ \, and \,
$\mu^+ \to e^+ \, \nu_e \, {\overline{\nu}}_{\mu}$). Neutrino oscillations
have been observed.$^{10}$ Further
exploration at a neutrino factory could reveal CP
violation in the lepton sector,$^{11}$ and is  particularly useful if the
coupling between $\nu_e$ and $\nu_{\tau}$, $\theta_{13}$, is small.$^8$  
A muon collider can do s-channel scans to separate 
the $H^0$ and $A^0$ Higgs doublet.$^{12}$ 
Above the ILC's 800 GeV there are a large array of supersymmetric particles
that might be produced,$^{13}$ as well as mini black holes,$^{14}$ if
large extra dimensions exist.  Note that the energy resolution of a muon 
collider is not smeared by beamstrahlung.  

\begin{figure}[t]
\centerline{\psfig{file=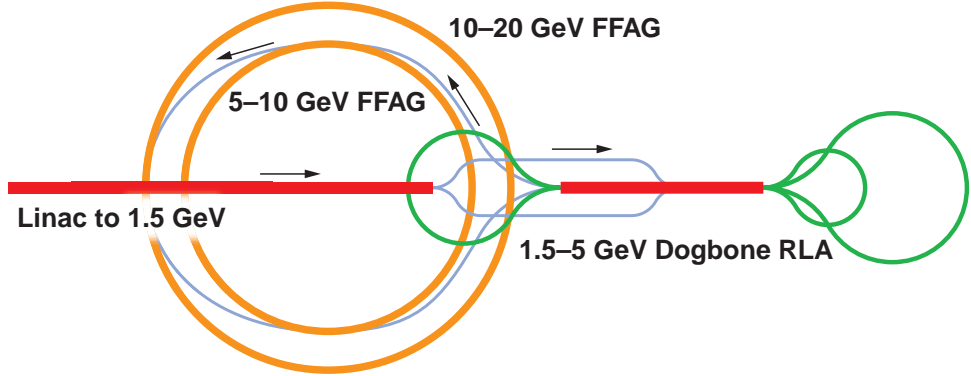,width=12.8cm}}
%\vspace*{8pt}
\caption{Possible 20 GeV muon accelerator layout from the Ref. 8 
neutrino factory design.}
\end{figure}

   A cyclotron has a large volume magnetic field which is constant in time.
Particle orbits move from the center to the edge of the cyclotron as they
accelerate.  A synchrotron has a small magnetic field volume. The $B$ field
increases with time.  Particle orbit radii do not change as a particle
accelerates. An FFAG ring is in between a cyclotron and a synchrotron in its
design. As particles accelerate they move a small distance in gradient magnets
which can accommodate higher energy orbits at slightly different radii. FFAG
magnetic fields are fixed in time and their volume is larger than a synchrotron
but smaller than a cyclotron.

\section{A Neutrino Factory Design using Two Non-Scaling FFAG Rings}

The most recent neutrino factory design$^8$ incorporates 5 $\to$ 10 and
10 $\to$ 20 GeV non-scaling FFAG rings. Acceleration up to 5 GeV uses
a linac and a dogbone recycling linac.$^{8,15}$  
A layout appears in Fig.~1 and parameters
in Table 1.  
The 20 GeV ring is almost five times larger than a synchrotron 
with 5.5 T magnets. The ratio of focusing--to--bending in an FFAG ring is high. 
Normally resonances are a problem in
non-scaling FFAGs, but the fast muon acceleration cycle can prevent
them from building up as can highly symmetric lattice designs. 
Each cell uses a FDF triplet of superconducting
magnets as shown in Fig.~2. Much work has gone into the lattice design 
to keep the beam size and hence the magnetic apertures  relatively small.
The idea is to control cost by reducing the magnetic field volume and
by using superconducting magnets with moderate fields.

Superconducting RF (fixed 201 MHz, 10 MV/m) is used for acceleration. A
niobium coated copper cavity running at 201 MHz has recently achieved
a gradient of 11 MV/m and prototypes may reach 15 MV/m.$^{16}$
At 201 MHz, ${1\over{4}} \lambda$ = 37 cm, on the same order as the phase
difference just due to the muons increasing in speed as shown in the last row
of Table 1.  
Its hard to change the RF phase itself quickly.
An advantage of the non-scaling FFAG is the additional control 
over the physical path length muons follow. Path lengths dominate speed
increases in determining muon phase with respect to the RF.
Fig. 3 notes the parabolic time of flight (TOF) 
relation that can be achieved. Muons cross the RF crest three times 
during the acceleration cycle.  Staying closer to crest minimizes the amount
and cost of RF that is needed. In a scaling FFAG, TOF increases monotonically.

\begin{table}[t]
\tbl{Neutrino Factory FFAG parameters. The LHC packing fraction is higher.} 
{\begin{tabular}{@{}lccc@{}} \toprule
              & Low Energy Ring & High Energy Ring & CERN LHC \\ \colrule
Ring Type & non-scaling FFAG & non-scaling FFAG & ramping synchrotron \\
Accelerated Particle & muon & muon & proton \\
Energy Range & 5 $\to$ 10 GeV & 10 $\to$ 20 GeV & 0.45 $\to$ 7 TeV \\
Ring Circumference & 400 m & 500 m & 27 km \\
Ring Radius (R)  & 64 m  & 80 m  & 4300 m \\ 
$B = p_{max} / .3 R$    & 0.52 T  & 0.83 T  & 5.4 T \\ 
$B_{max}$           & 4.2 T  & 5.5 T  & 8.4 T \\ 
Magnet Packing Fraction  & 12\%  & 15\%  & 64\% \\ 
RF Characteristics & 10 MV/m, 201 MHz & 10 MV/m, 201 MHz & 200 $\to$ 400 MHz \\ 
RF Energy Extracted  & 16\%  & 27\%  &  \\ 
Total RF Voltage  & 480 MV  & 578 MV  &  \\ 
Initial Speed ($\beta = p/E$) & 0.999777   & 0.999944   & 0.999997826 \\
Final Speed   ($\beta = p/E$) & 0.999944   & 0.999986   & 0.999999991 \\
Orbits to $E_{max}$  & 9.6  & 16.5  & 13 million  \\ 
Acceleration Time & 13 $\mu$S  & 28 $\mu$S   & 20 minutes \\
Particle Decay Loss  & 9\%  & 10\%  &  0\% \\ 
$c(\beta_f - \beta_i)({\rm Time})/2$ &  32 cm  & 17 cm  & 380 km \\ \botrule 
\end{tabular}}
\end{table}

\begin{figure}[b]
\begin{tabular}{cc}
\begin{minipage}[t]{6cm}
\centerline{\psfig{file=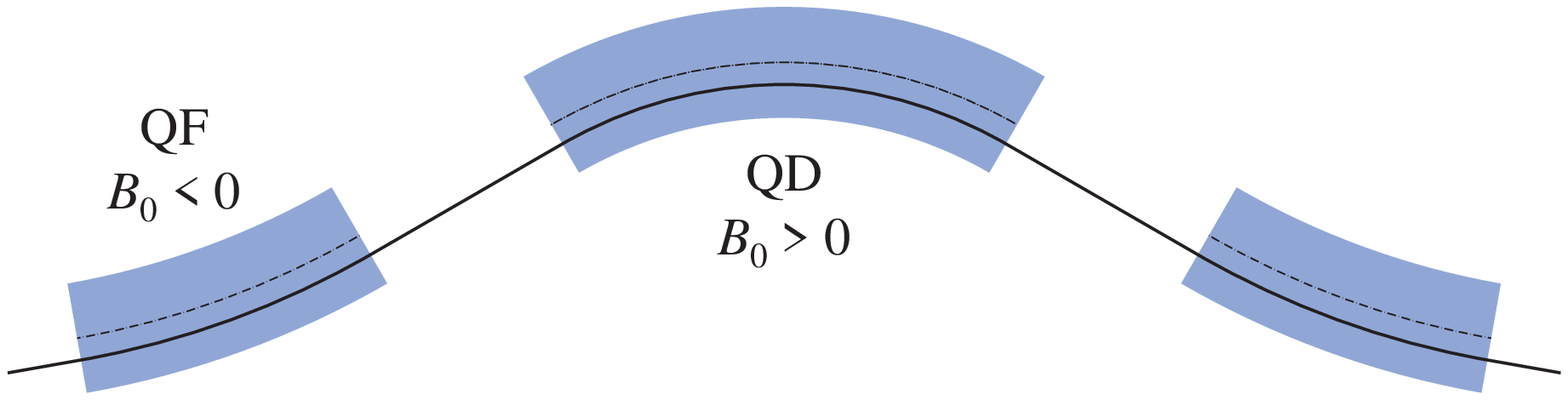,width=6cm}}
%\vspace*{8pt}
\caption{Triplet 
Focusing--Defocusing--Focus\-ing (FDF) 
lattice geometry for the superconducting magnets in 
the 5--10 and 10--20 GeV non-scaling 
FFAG neutrino factory rings.$^8$}
\end{minipage}

&

\begin{minipage}[t]{6cm}
\centerline{\psfig{file=tof.eps,width=6cm}}
%\vspace*{8pt}
\caption{Time of flight difference per magnet cell vs.~energy for the
5 to 10 GeV non-scaling FFAG ring in the Ref. 8  neutrino factory design. 
Muons must stay in phase with the RF.}
\end{minipage}
\\
\end{tabular}
\end{figure}

\section{Scaling FFAG Rings Being Built in Japan}
\vspace*{-2pt}

FFAGs are being built for muon phase rotation, radiation therapy, CT scanning,
and accelerator--driven sub--critical nuclear reactor operation in Japan. 
A 1 MeV scaling FFAG with 8 DFD sectors 
has been built at KEK
and has accelerated protons with a 1 kHz repetition 
rate.$^{5,17}$ 
A 150 MeV scaling FFAG with 12 DFD sectors is
nearing completion. Beam has been accelerated to 150 MeV. 
Orbits shift 
from a radius of 4.4 to 5.5 m during the acceleration cycle.
In these scaling FFAGs, orbit shapes and
magnet focal lengths are energy independent. See Fig. 2
of Ref.~5 for a nice drawing of particle paths in scaling and
non-scaling FFAGs.  

\vspace*{-6pt}
\section*{Acknowledgments}
\vspace*{-2pt}

Many thanks to J.~Gallardo, S.~Berg, C.~Johnstone, 
R.~Palmer, and Y.~Torun.

\end{document}